# The quest for explosive bubbles in the Indonesian Rupiah/US exchange rate: Does the uncertainty trinity matter?

## Abdul Khaliq[a*], Syafruddin Karimi[a], Werry Darta Taifur[a], Endrizal Ridwan[a]

[a]Department of Economics, Universitas Andalas, Padang, Indonesia

| CHRONICLE | ABSTRACT |
|---|---|
|  | The Generalized Supremum Augmented Dickey-Fuller (GSADF) technique is performed to resolve whether the Indonesian Rupiah/US exchange rate has experienced multiple explosive bubbles. The GSADF uncovers that the Indonesian Rupiah/US exchange rate deviates from the fundamental values by six times from January 1985 to September 2023, periodically indicating the presence of numerous explosive behaviors. Once the full-sample period separates into the managed-floating regime and the free-floating regime, the GSADF still detects multiple bubbles. Of particular curiosity on uncertainty trinity, this study underlines that global geopolitical risk negatively drives explosive actions in the ratio of exchange rates for non-traded and traded goods. The global economic policy uncertainty negatively affects speculative bubbles in the exchange rate and the ratio of exchange rates for non-traded. The country's geopolitical risks negatively strike only speculative bubbles in the exchange rate. Further, we find heterogeneity in our results by examining different exchange rate systems. The robustness checks further firmly ascertain across baseline empirical findings.<br><br>© 2024 by the authors; licensee Growing Science, Canada. |

## 1. Introduction

This study has two main touch insights. The first sight is to quest for explosive performance episodes in the Indonesian Rupiah against the US exchange rate by employing Phillips et al. (2015a) generalized sup ADF (GSADF) test. The second insight is to feature the force of global geopolitical risk and economic policy uncertainty together with country-specific geopolitical risk, which Caldara and Iacoviello (2022) termed the "uncertainty trinity," to shed light on explosive actions in the Indonesian Rupiah-US dollar. The exchange rate dynamic is a crucial subject that dictates the value of options, making investment decisions, and hedging options. Therefore, it has become a significant theme of discussion among academics, regulators, and entrepreneurs in the financial markets and assets (Caporale et al., 2015; Salisu, 2020). Most recent produces of papers on the exchange rate extensively analyze the fluctuation of exchange rate unpredictability (among others, Aftab et al., 2023; Bush & Noria, 2021; Luo et al., 2023; Pastorek, 2023; Peng et al., 2023; Aysun, 2024) and the explosive activities in the exchange rates (Jirasakuldech et al., 2006; Bettendorf & Chen, 2013; Jiang et al., 2015; Hu & Oxley, 2017; Ural, 2021; Yildirim et al., 2022).

To date, detecting the speculative action in the exchange rate has become debatable and continues to be mystifying. The inquiry remains about the sources of the explosive bubbles in exchange rates mainly due to global- and country-specific uncertainty. We argue that global- and country-specific uncertainty are the root causes of speculative bubbles as identified in the drivers of exchange rate instability studied by, for instance, Pástor and Veronesi (2013), Bartsch (2019), Chen et al. (2020), Bush and Noria (2021), Khaliq (2022), and Salisu, et al. (2022). Pástor and Veronesi (2013) are concerned about political uncertainty. Khaliq (2022) and Salisu et al. (2022) feature geopolitical risks. Bartsch (2019) and Chen et al. (2020) spotlight the role of economic policy uncertainty. We also underline that the influence of global- and country-specific uncertainty on the rational speculative bubbles has the same logic as on exchange rate dynamic, that is, through the financial markets and assets (Chiang, 2021; Hoque et al., 2021; Iyke et al., 2022; Hossain et al., 2023).

However, until now, there have been very few academic works attempting to detect the particular explosive bubbles of the exchange rate movement between the Indonesian Rupiah contrary to the US Dollar (e.g., Hu & Oxley, 2017). Indonesia's exchange rate system has been changed over time. From a managed floating exchange rate that was in place until July 1997 to the operation of a free-floating exchange rate policy after July 1997. We hypothesize that the exchange rate system changes will direct rational speculative bubbles' existence. We claim that, prior to this study, there has been a notable absence of substantial works to deliver explicit issues encompassing the sources of explosive behavior in the Indonesian Rupiah against the US Dollar. In light of these insights, Indonesia appears to be an attractive and thought-inspiring place for the investigation of explosive bubbles. Hence, this study can provide novelty to this research gap.

Our empirical construction is executing the sup ADF (SADF) and generalized sup ADF (GSADF) to uncover explosive behavior. When the explosive behavior is detected, we apply logit regression to seek the upshot of global uncertainty and specific country geopolitical risks on the explosive behavior. The novelty idea to investigate the global uncertainty and specific country geopolitical risks on the explosive bubbles of the Rupiah as opposed to the US dollar exchange rate is due to, between January 1985 and September 2023, 1) the circumstances of global uncertainty and country-specific geopolitical risks are pretty demanding, such as in specific reforming governmental institutions in Indonesia from the authoritarian regime to democracy, terror attacks, the global geopolitical risks such as Russian-Ukraine war, Israel-Palestinian war, etc., 2) the altering of Indonesian exchange rate regime from the managed-floating regime (until July 1997) to the free-floating regime (after July 1997-now), 3) a variety of financial crises either Global Financial Crisis (GFC) or Asian Financial Crisis (AFC), and 4) The COVID-19 pandemic hit Indonesian from March 2020 to June 2023.

## 2. Literature Review

Currently, much academic research on financial markets and assets has been focusing on detecting portfolios of asset price bubbles. Albeit various papers with comprehensive topics and diverse approaches surveyed explosive bubbles, the well-known methods used are SADF and GSADF (Phillips et al., 2011; Phillips & Shi, 2020; Phillips et al., 2015a, 2015b; Hu, 2023). The SADF and GSADF are successful research methods for searching bubbles in stock markets, precious metal markets, cryptocurrency markets, housing markets, and foreign exchange rate markets.

Several recent studies employing the SADF and GSADF to analyze stock markets were conducted, e.g., Chen et al. (2015), Caspi and Graham (2018), El Montasser et al. (2018), Zhang et al. (2020), and Potrykus (2023). The applications of SADF and GSADF are also realized in the precious metal markets (Su et al., 2017; Khan & Köseoğlu, 2020; Ma & Xiong, 2021; Ozgur et al., 2021; Wang et al., 2023). The SADF and GSADF are commonly directed at detecting bubble behavior in the commodity markets (Huang & Xiong, 2020; Wang et al., 2020; Khan et al., 2021; Lawal et al., 2022; Akcora & Kocaaslan, 2023; Fang et al., 2023) and housing markets (Huang and Shen, 2017; Tsai and Chiang, 2019; Coskun et al., 2020; Martínez-García & Grossman, 2020; Chen et al., 2021; Tsai & Lin, 2022; Trojanek et al., 2023). Further, the SADF and GSADF are intensively instigated in cryptocurrency markets (Cheung et al., 2015; Bouri et al., 2019; Bazán-Palomino, 2022; Gemici et al., 2023; Guo et al., 2023; Haykir & Yagli, 2022; Chowdhury & Damianov, 2024) and foreign exchange rate markets (El Montasser et al., 2016; Hu & Oxley, 2017; Maldonado et al., 2021; Ural, 2021; Yildirim et al., 2022).

Although many examples of published papers discuss utilizing GSDAF to uncover the presence of explosive behavior, only a few pieces of systematic literature spotlight the sources of speculative bubbles, see, i.e., Khan et al. (2021), Su et al. (2023), and Wang et al. (2023). In particular, in the studies of exchange rates (see, i.e., Jirasakuldech et al., 2006; Hu & Oxley, 2017; Ural, 2021; Yildirim et al., 2022), to the finest of our expertise, lack of published papers empirically standpoint on and shape the sources of those explosive bubbles that regulators and investors are concerned about, especially related to the uncertainty trinity. A particular study on Indonesia's Rupiah against the US exchange rate was conducted by Hu and Oxley (2017); they only detected multiple bubbles without splitting up exchange rate regimes and did not question the sources of the bubbles. Hence, this paper serves as an academic reference to fill these research gaps.

## 3. Theoretical Model

In line with Engel and West (2005), Bettendorf and Chen (2013), León-Ledesma and Mihailov (2014), Jiang et al. (2015), and Hu and Oxley (2017), we assume the model of the exchange rate is strong-minded by actual and estimated values of fundamentals as follows

$$s_t = (1 - \alpha) \sum_{j=0}^{k} \alpha^j E_t(f_{t+j}) + \alpha^{k+1} E_t(s_{t+k+1})$$

(1)

where $s_t$ represents the nominal exchange rate at time $t$, $f_t$ stands for the market fundamental at time $t$, and $\alpha$ is the discount factor. In long-run, we presume the exchange rate is only determined by future expected fundamentals, therefore,

$$\lim_{k \to \infty} \alpha^{k+1} E_t(s_{t+1})$$

(2)

Yet, the explosive behavior of the exchange rate may emerge if the transversality condition does not handhold following an autoregressive (AR(1)) process,

$$\eta_t = \frac{1}{\alpha}\eta_{t-1} + \varepsilon_t \tag{3}$$

where $\frac{1}{\alpha}$ is greater than 1, as an explosive behavior process, we, then, state the exchange rate as

$$s_t = s_t^f + \eta_t \text{ or } s_t^f = s_t - \eta_t \tag{4}$$

where $s_t^f$ represents future exchange rate fundamental values and $\eta_t$ signifies the bubble factor. Following Engel and West (2005), $f_t$ is assumed to be I(1). Based on purchasing power parity (PPP) framework, the price-differential is the basis for the nominal exchange rate fundamental:

$$f_t = p_t - p_t^* \tag{5}$$

where $p_t$ and $p_t^*$ are domestic and foreign prices, respectively. Following Engel (1999), the price index could be classified as below

$$p_t = (1 - \beta)p_t^T + \beta p_t^N \tag{6}$$

where $p_t^T$ and $p_t^N$ are the logarithmic price indices of traded and non-traded goods component, correspondingly. In parallel idea, the overseas price index could be expressed as

$$p_t^* = (1 - \gamma)p_t^{T*} + \gamma p_t^{N*} \tag{7}$$

The price differential based on Eq. (6) and Eq. (7) can be derived as follows

$$p_t - p_t^* = (p_t^T - p_t^{T*}) + \beta(p_t^N - p_t^T) - \gamma(p_t^{N*} - p_t^{T*}) \tag{8}$$

The producer price index (PPI) is implemented here as proxied the price level traded goods component respecting Engel (1999):

$$f_t^T = p_t^T - p_t^{T*} = \ln(PPI_t) - \ln(PPI_t^*) \tag{9}$$

The comparative consumer price indices (CPI) to producer price index (PPI) measures non-traded goods component:

$$\begin{aligned} f_t^N &= \beta(p_t^N - p_t^T) - \gamma(p_t^{N*} - p_t^{T*}) \\ &= \ln(CPI_t) - \ln(PPI_t) - \left(\ln(CPI_t^*) - \ln(PPI_t^*)\right) \end{aligned} \tag{10}$$

## 4. Estimation Method and Data

### 4.1. Estimation Method

To capture the explosive behavior, we employ a sup ADF (SADF) test proposed by Phillips et al. (2011). Homm and Breitung (2012) empirically prove the ability of SADF to detect the existence of explosive bubbles. However, the SADF test fails to uncover the presence of multiple bubbles (Phillips et al., 2015). The test technique is grounded on the time-varying autoregressive specification

$$\Delta s_t = \mu_{r_1,r_2} + \delta_{r_1,r_2}s_{t-1} + \sum_{i=1}^{k}\phi_{r_1,r_2}^i \Delta s_{t-1} + \varepsilon_t, \quad \varepsilon_t \sim N\left(0, \sigma_{r_1,r_2}^2\right) \tag{11}$$

where $s_{t-1}$ is the logarithmic Indonesian Rupiah/US exchange rate, and $k$ signifies the lags number. In the recursive unit root test, the null hypothesis is $H_0: \delta = 1$ and $H_1: \delta > 1$, of Eq. (11) is regressed repetitively treating sample data augmented by one observation at every try. The outcome of statistics is by the subsequent equation:

$$ADF_r \Rightarrow \frac{\int_0^r \widetilde{W}\, dW}{\left(\int_0^r \widetilde{W}^2\right)^{1/2}} \tag{12}$$

with the boundary distribution:

$$\sup_{r\in[r_0,1]} ADF_r \Rightarrow \sup_{r\in[r_0,1]} \frac{\int_0^r \widetilde{W}\, dW}{\left(\int_0^r \widetilde{W}^2\right)^{1/2}} \tag{13}$$

The window size $r_w$ magnifies from $r_0$ to 1. The initial point $r_1$ is set at zero, and the terminal date of each sample ($r_2$) is equivalent to $r_w$ and adjusts from $r_0$ to 1. The statistic sequence of SADF is therefore delineated as below

$$SADF(r_0) = \sup_{r_2 \in [r_0, 1]} ADF_0^{r_2}$$

(14)

Since SADF has constraint in window size, the technique is then developed to GSADF (Phillips et al., 2015), which has flexible window size in inquiring existence of the multiple bubbles. The terminal point $r_2$ alters from $r_0$ to 1, where the initial point $r_1$ is also permitted from 0 to $r_2 - r_0$. The important variance between SADF and GSADF is the window size of the initial point $r_1$. The statistic of GSADF is presented as follows

$$GSADF(r_0) = \sup_{\substack{r_2 \in [r_0, 1] \\ r_1 \in [0, r_2 - r_0]}} ADF_{r_1}^{r_2}$$

(15)

The GSADF statistics test has a limit distribution when considering unsystematic walk intercept and the null hypothesis in the estimation method as;

$$\sup_{\substack{r_2 \in [r_0, 1] \\ r_1 \in [0, r_2 - r_0]}} \left\{ \frac{\frac{1}{2} r_w [W(r_2)^2 - W(r_1)^2 - r_w] - \int_{r_1}^{r_2} W(r) dr [W(r_2) - W(r_1)]}{r_w^{1/2} \left\{ r_w \int_{r_1}^{r_2} W(r)^2 dr - \left[ \int_{r_1}^{r_2} W(r) dr \right]^2 \right\}^{1/2}} \right\}$$

(16)

where $r_w = r_2 - r_1$ and $W$ is a typical Wiener process

Thus, the approach of SADF and GSADF to detect explosive behavior is established on the backward sup ADF (BSADF) test as follows

$$BSADF_{r_2}(r_0) = \sup_{r_1 \in [0, r_2 - r_0]} ADF_{r_1}^{r_2}$$

(17)

Therefore, the j-th bubble's initial and terminal dates are given by;

$$\tilde{r}_{js} = \inf_{r_{2 \in (r_0, 1)}} \{r_2 : BSADF_{r_2}(r_0) > cv_{r_2}^{\alpha T}\}$$

(18)

$$\tilde{r}_{je} = \inf_{r_{2 \in \left( \tilde{r}_{je} + \delta log(T)/T, 1 \right)}} \{r_2 : BSADF_{r_2}(r_0) < cv_{r_2}^{\alpha T}\}$$

(19)

where $r_e$ and $r_f$ are the initial and end points of a bubble, separately. The $cv_{r_2}^{\alpha T}$ is the critical value of $100(1 - \alpha)\%$ for $[r_2 T]$.

This analysis estimates the occurrence of multiple bubbles and establishes the initial and end points of the bubbles. Likewise, the report investigates the important role of global geopolitical risk and economic policy uncertainty as well as country geopolitical risks, known as uncertainty trinity, in exchange rate bubbles which is outlined by $R_t$ as;

$$R_t = \begin{cases} 0, if \ BSADF_t(r_0) < cv_{r_2}^{\alpha T} \\ 1, , if \ BSADF_t(r_0) > cv_{r_2}^{\alpha T} \end{cases}$$

(20)

When explosive behavior in Indonesian Rupiah/US is spotted then $R_t$ is equal to 1 and otherwise. The sources of explosive bubbles from uncertainty trinity are tested through logit regression. We construct as follows

$$R_t = A_t' \alpha + \mu_t$$

(21)

where $A_t$ is the uncertainty trinity factors on explosive bubbles. The logit model is styled as follows

$$(R_t | A_t) = \varphi(A_t' \alpha)$$

(22)

The log-likelihood function estimates the cause-effect mean parameters and is constructed as below

$$lnL = \sum_{t=1}^{T} R_t ln[\varphi(A_t' \alpha)] + \sum_{t=1}^{T} (1 - R_t) ln[1 - \varphi(A_t' \alpha)]$$

(23)

The degree of influence of the uncertainty trinity factors on explosive bubbles is measured by the marginal effect:

$$\frac{\partial P(R_t = 1 | A_t)}{A_j} = \varphi(A_t' \alpha) . \alpha_j \tag{24}$$

### 4.2. Data

We examine the explosive bubbles of the Indonesian Rupiah compared to the US dollar exchange rate from January 1985 (1985M1) through September 2023 (2023M9), published in https://www.bi.go.id/. We classify into three periods of data: the period of the managed-floating regime (1985M1:1997M7), the period of the free-floating rule (1997M8:2023M9), and, finally, the full-sample range (1985M1:2023M9). This study also utilizes the Indonesian CPI, the US CPI*, the Indonesian producer price indices (PPI), and the US producer price indices (PPI*) served by subscribed https://eikon.refinitiv.com/. The global uncertainty data are proxied by global geopolitical risk (GPR) and the US economic policy uncertainty (GEPU). The US monetary policy uncertainty (GMPU) is considered for robustness checks purposes. The geopolitical risks data is taken from https://www.matteoiacoviello.com/gpr.htm, while the US economic policy uncertainty and the US monetary policy uncertainty are public access from https://www.policyuncertainty.com/.

## 5. Results and Discussion

### 5.1. Baseline Estimation

Table 1 illustrates the estimation results of ADF, SADF, and GSADF to inquire about explosive bubbles in the Indonesian Rupiah against the US exchange rates. Each result is derived by classifying three scenarios, that is, full sample, managed-floating regime, and free-floating regime, with initial window sizes of 23, 35, and 43 observations, respectively. For each scenario, the SADF is significant at a 1% level for the full sample scenario only, whilst the GSADF is statistically significant for all scenarios except the ratio exchange rate to traded goods component at the managed-floating regime. The GSADF uncovers that the Indonesian Rupiah/US exchange rate deviates from the fundamental values by six times from January 1985 to September 2023, which directs the occurrence of numerous explosive behaviors. When the full-sample period distinguishes between the managed-floating regime for the period January 1985 to July 1997 and the free-floating regime in the period July 1997 to September 2023, the GSADF still observes the presence of multiple bubbles at 1% and 5% of the thresholds, disparately.

**Table 1**
The Results of Detecting Explosive Bubbles Using ADF, SADF, and GSADF Tests for Indonesian Rupiah/US Exchange Rate

| Variables | Full Sample 1985M1-2023M9 | | | |
|---|---|---|---|---|
| | ADF | SADF | GSADF | Episodes |
| $s_t$ | -1.6618 | 2.9126*** | 6.3442*** | 90M04-1993M09, 1994M02, 1995M07-1996M08, 1997M03-1998M08, 2013M09-2013M12, 2015M09 |
| $s_t - f_t^N$ | -1.1692 | 2.4161*** | 5.3734*** | 1990M09-1990M11, 1995M12, 1996M03-1996M12, 1997M01, 1997M07-1998M07, 2012M09-2014M02, 2014M06, 2018M10, 2022M05-2022M07 |
| $s_t - f_t^T$ | -2.6462 | 2.5137*** | 5.7778*** | 1997M12, 1998M01-1998M02, 1998M06, 2013M09, 2013M12 |
| CV 1% | 0.8114 | 2.0396 | 2.6111 | |
| CV 5% | -0.0050 | 1.4818 | 2.1957 | |
| CV 10% | -0.3824 | 1.1881 | 1.9464 | |
| | Managed Floating Regime: 1985M1-1997M7 | | | |
| $s_t$ | -1.7844 | -0.3833 | 3.4503*** | 1988M09-1993M09, 1995M06-1996M05, 1998M09, 1997M04, 1997M06-1997M07 |
| $s_t - f_t^N$ | -1.5605 | -0.3762 | 4.2790*** | 1989M01-1989M07, 1990M01-1990M03, 1990M09-1990M10, 1995M01-1995M06, 1995M12, 1996M03-1997M01, 1997M07 |
| $s_t - f_t^T$ | -2.7506 | -0.6461 | 0.7579 | NEB |
| CV 1% | 0.6379 | 1.9305 | 2.6739 | |
| CV 5% | 0.0471 | 1.3618 | 2.1128 | |
| CV 10% | -0.4511 | 1.0483 | 1.8131 | |
| | Free Floating Regime: 1997M8-2023M9 | | | |
| $s_t$ | -4.7460 | -3.2857 | 2.4666** | 2013M08-2014M01, 2015M09 |
| $s_t - f_t^N$ | -2.4982 | -2.4881 | 2.5161** | 2012M05, 2013M08-2014M02, 2014M06, 2018M09-2018M10, 2022M05-2022M07, 2022M09-2022M12 |
| $s_t - f_t^T$ | -2.1084 | -1.5310 | 2.4882** | 2013M09-2014M01 |
| CV 1% | 0.6828 | 1.9994 | 2.5712 | |
| CV 5% | -0.0296 | 1.4113 | 2.1141 | |
| CV 10% | -0.4280 | 1.1170 | 1.8914 | |

Note: CV states critical values developed by Vasilopoulos et al. (2020). The initial window size for the managed floating regime is 23 observations, the free-floating regime is 35 observations, and the full sample is 43 observations for ADF, SADF, and GSADF tests. NEB stands for no explosive bubbles found. ***, **, * are significant at 1%, 5%, and 10%, correspondingly.
Source: Authors' calculation

Fig. 1 graphically presents the date-stamping analysis of detecting the explosive bubbles utilizing ADF, SADF, and GSADF checks for the Indonesian Rupiah/US Exchange Rate. In the full sample data, we show six times of explosive bubble episodes in $s_t$, including 1990M04-1993M09, 1994M02, 1995M07-1996M08, 1997M03-1998M08, 2013M09-2013M12,

and 2015M09 from Fig.1a. The longest bubble episode is 42 months. When the full sample data is tested to sub-samples of $s_t - f_t^N$ and $s_t - f_t^T$, we found different multiples bubbles. The $s_t - f_t^N$ presents nine times of explosive behavior from Fig. 1b, comprising 1990M09-1990M11, 1995M12, 1996M03-1996M12, 1997M01, 1997M07-1998M07, 2012M09-2014M02, 2014M06, 2018M10, and 2022M05-2022M07. The $s_t - f_t^T$ displays five times speculative bubbles from Fig. 1c, containing 1997M12, 1998M01-1998M02, 1998M06, 2013M09, and 2013M12. The first through the third bubble episodes of $s_t$ no longer emerge if the $f_t^T$ is factored into the model, whilst the sixth episode of the bubble disappears from $s_t$, once $f_t^N$ is considered.

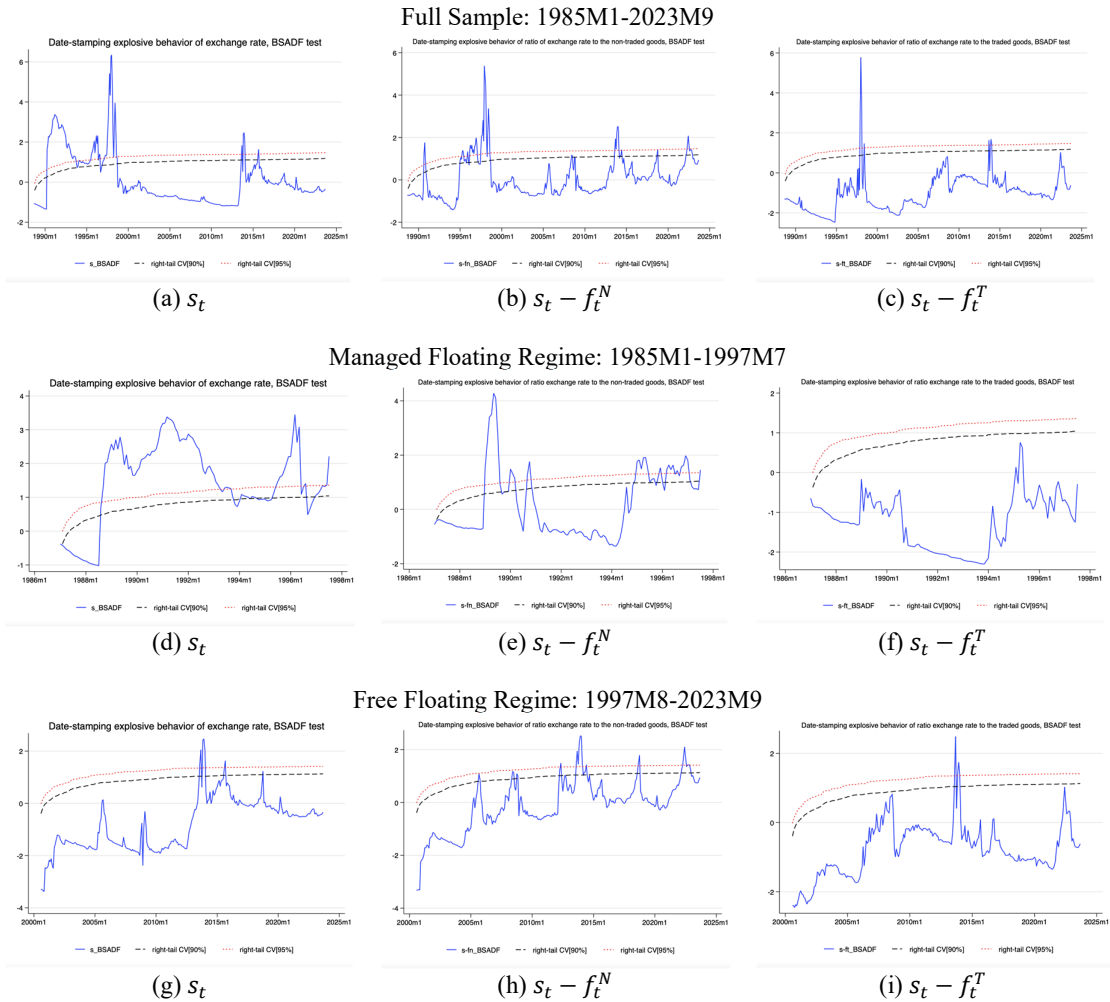

Source: Authors' calculation

**Fig.1.** Date-stamping analysis for the exchange rate ($s_t$), the ratio exchange rate to the Non-traded Goods ($s_t - f_t^N$), and the ratio exchange rate to the traded Goods ($s_t - f_t^T$)

Fig 1. d-f presents the explosive bubbles in time of the managed floating exchange rate regime. we reveal six times of explosive bubble episodes in $s_t$, including 1988M09-1993M09, 1995M06-1996M05, 1998M08, 1997M04, and 1997M06-1997M07 from Fig.1d. The longest bubble episode is 60 months. Whilst the full sample data is subtracted to sub-samples, that is $s_t - f_t^N$ and $s_t - f_t^T$, we uncovered dissimilar multiples bubbles. The $s_t - f_t^N$ gives seven times of explosive behavior from Fig. 1e, encompassing 1989M01-1989M07, 1990M01-1990M03, 1990M09-1990M10, 1995M01-1995M06, 1995M12, 1996M03-1997M01, and 1997M07. The multiple bubble episodes in $s_t$ over the period 1991M01-1993M09 no longer existed. Extremely, the explosive bubbles in the $s_t - f_t^T$ totally disappear from Fig. 1f. This finding implies that the $f_t^T$ has a significant effect on the bubbles in $s_t$ in time of managed floating exchange rate regime.

Fig 1. g-i displays the explosive bubbles in the period of the free-floating exchange rate system. we expose only two times of explosive bubble episodes in $s_t$, including 2013M08-2014M01 and 2015M09 from Fig.1g. Once the full sample data is examined to sub-samples of $s_t - f_t^N$ and $s_t - f_t^T$, we clarified distinctive multiples bubbles. The $s_t - f_t^N$ shows six times explosive behavior from Fig. 1h, comprising 2012M05, 2013M08-2014M02, 2014M06, 2018M09-2018M10, 2022M05-2022M07, and 2022M09-2022M12. The $s_t - f_t^T$ displays only one-time speculative bubbles from Fig. 1i, which is 2013M09-2014M01. Interestingly, the free-floating system leads to a short period of bubbles, with the longest bubble being 8 months. The bubbles episode in the free-floating system shrinks seven times compared to the managed floating exchange rate system.

**Table 2**
The Estimation Results of the Influence of Uncertainty Trinity on Explosive Bubbles

| Explanatory Variables | Full-Sample Range: 1985M1:2023M9 | | | | | |
|---|---|---|---|---|---|---|
| | $s_t$ | | $s_t - f_t^N$ | | $s_t - f_t^T$ | |
| | Coefficients | Marginal Effect | Coefficients | Marginal Effect | Coefficients | Marginal Effect |
| *GPR* | -0.4366 | -0.0560 | -2.3036*** | -0.1430*** | -3.8810*** | -0.0257** |
| | (0.4614) | (0.0589) | (0.6055) | (0.0349) | (1.4968) | (0.0112) |
| *GEPU* | -1.1098*** | -0.1423*** | -1.1843** | -0.0735** | 0.2010 | 0.0013 |
| | (0.3595) | (0.0442) | (0.4861) | (0.0288) | (0.9867) | (0.0065) |
| *GPRI* | -9.7937** | -1.2560** | 2.2768 | 0.1414 | -1.7186 | -0.0114 |
| | (4.5280) | (0.5503) | (3.1959) | (0.2002) | (13.4547) | (0.0881) |
| constant | 5.9575** | | 13.3304*** | | 11.7669 | |
| | (2.5053) | | (3.3063) | | (7.1874) | |
| Log likelihood | -202.2301 | | -125.9779 | | -28.4285 | |
| LR Statistics | 19.3600 | | 25.4500 | | 7.2700 | |
| Prob>Chi square | 0.0002 | | 0.0917 | | 0.0638 | |
| Observations | 465 | | 465 | | 465 | |
| McFadden's $R^2$ | 0.0460 | | 0.0920 | | 0.1130 | |
| | Managed Floating Regime: 1997M8-2023M9 | | | | | |
| *GPR* | -0.2376 | -0.0594 | -2.0696** | -0.2712** | | |
| | (0.6246) | (0.1562) | (0.9669) | (0.1218) | | |
| *GEPU* | 0.8158 | 0.2039 | -2.4010** | -0.3147** | | |
| | (0.7232) | (0.1808) | (1.0480) | (0.1297) | | |
| *GPRI* | -2.4492 | -0.6123 | 3.8761 | 0.5080 | | |
| | (6.9392) | (1.7347) | (9.4797) | (1.2599) | | |
| constant | -2.5624 | | 18.5606*** | | | |
| | (3.3259) | | (5.6949) | | | |
| Log likelihood | -103.9321 | | -65.5950 | | | |
| LR Statistics | 1.4600 | | 16.5400 | | | |
| Prob>Chi square | 0.6916 | | 0.0009 | | | |
| Observations | 151 | | 151 | | | |
| McFadden's $R^2$ | 0.0070 | | 0.1120 | | | |
| | Free Floating Regime: 1997M8-2023M9 | | | | | |
| *GPR* | 0.2784 | 0.0005 | 0.7089 | 0.0302 | -0.7912 | -0.0004 |
| | (1.9412) | (0.0035) | (0.9192) | (0.0378) | (2.3222) | (0.0010) |
| *GEPU* | -1.1298 | -0.0020 | 0.6442 | 0.0274 | -1.6997 | -0.0009 |
| | (1.2360) | (0.0029) | (0.6104) | (0.0256) | (1.5928) | (0.0017) |
| *GPRI* | -86.7476** | -0.1524 | -10.8703 | -0.4624 | -102.3001** | -0.0548 |
| | (35.0426) | (0.1565) | (9.1226) | (0.3525) | (47.0654) | (0.0871) |
| constant | 2.5686 | | -8.7962* | | 9.8335 | |
| | (10.5284) | | (4.8953) | | (12.4031) | |
| Log likelihood | -24.0805 | | -61.4341 | | -16.7672 | |
| LR Statistics | 11.2200 | | 3.5600 | | 9.3200 | |
| Prob>Chi square | 0.0106 | | 0.3131 | | 0.0253 | |
| Observations | 314 | | 314 | | 314 | |
| McFadden's $R^2$ | 0.1890 | | 0.0280 | | 0.2170 | |

Note: ***, **, * are significant at 1%, 5%, and 10%, separately.
Source: Authors' calculation

Although we shed light on the presence of multiple bubbles episodes in the Indonesia Rupiah against the US exchange rate, the question still remains on the role of uncertainty trinity plays. Referring to many recently published papers strongly captured the role of uncertainty on exchange rate impulsiveness; see among recent studies, Chen et al. (2020), Bush and Noria (2021), Salisu, et al. (2022), Singh et al. (2022), and Che et al. (2023). We then take into account the influence of the uncertainty trinity on the explosive bubble incidences. The estimation technique further employs the logistic regression in dealing with Eq. (21), see detailed estimation results in Table 2. Of particular questioning on uncertainty trinity with the full sample data (1985M1:2023M9), we highlight that global geopolitical risk negatively determines explosive bubbles in of $s_t - f_t^N$ and $s_t - f_t^T$ and is significant at a 1% level. This result is in line with Iyke et al. (2022) and Hossain et al. (2023), who concluded the negative effect of global geopolitical risks on the foreign exchange market. The marginal effect shows that a unit alteration in global geopolitical risk leads to a 14,3% decrease in $s_t$ and a 2,57% decrease in $s_t - f_t^T$. The Indonesia-specific geopolitical risks negatively discover explosive bubbles in $s_t$ only and are statistically significant at a 1% level (Iyke et al., 2022). Similarly, a unit change in a country's geopolitical risk precedes a 125% drop in the exchange rate bubbles. Furthermore, the global economic policy uncertainty negatively shakes speculative bubbles in $s_t$ and $s_t - f_t^N$ and are significant at 1% and 5% levels, respectively. These findings are coherent with Abdulsalam and Onipede (2023). A unit change in the global economic uncertainty directs to a worsening in $s_t$ and $s_t - f_t^N$ by 14,2% and 7,4%, respectively.

Intriguingly, we find heterogeneousness in our findings when breaking down the sample data by exchange rate scheme, that is the managed floating exchange rate system (1985M1;1997M7) and free-floating exchange rate regime (1997M8:2023M9). We find that the global geopolitical risk and the global economic policy uncertainty negatively impress speculative bubbles in $s_t - f_t^N$ only and are statistically significant at 5% levels in the managed floating exchange rate regime. These results are consistent with Che et al. (2023). The marginal effect presents a unit change in global geopolitical risk and global economic policy uncertainty dropping to 27,1% and 31,5% in $s_t - f_t^N$. However, in the free-floating exchange rate regime, the Indonesia-specific geopolitical risk negatively affects speculative bubbles in $s_t$ and $s_t - f_t^T$ and is statistically significant at 5% levels. The reinforcement of global- geopolitical risk and economic policy uncertainty has no longer persistence. A unit change in the Indonesia-specific geopolitical risk points to a weakening in the exchange rate and the ratio of exchange rates for traded goods by 15,2% and 5,5%, one-to-one. Yet, this finding contradicts Khaliq (2022), who states the positive consequence of Indonesia-specific geopolitical risk on exchange rate volatility. In fact, empirically, exchange rate volatility is different from explosive bubbles in the Rupiah-US dollar. The insinuation of this result is that the respective bubbles in exchange rate systems respond un-similarly to the uncertainty trinity, an occurrence sound as the Meese-Rogoff puzzle (Meese and Rogoff, 1983).

### 5.2. Robustness Checks

It is important to take into account that the empirical results that have been analyzed thus far may not always line up. We perform a robustness test to inspect the consistency of the outcomes. The historical geopolitical risk for the entire world (GRPH), the US monetary policy uncertainty (GMPU), and the historical geopolitical risk unique to Indonesia (GRPHI) are used as proxied variables for the robustness checking tests in our methodology. As a result, Table 3 displays the detailed robustness checking findings.

**Table 3**
Robustness Test Results of the Influence of Uncertainty Trinity on Explosive Bubbles

| Explanatory Variables | Full-Sample Range: 1985M1:2023M9 | | | | | |
| | $s_t$ | | $s_t - f_t^N$ | | $s_t - f_t^T$ | |
| | Coefficients | Marginal Effect | Coefficients | Marginal Effect | Coefficients | Marginal Effect |
| --- | --- | --- | --- | --- | --- | --- |
| *GPRH* | -0.2833 | -0.0330 | -2.1709*** | -0.1406*** | -3.3859*** | -0.0241** |
| | (0.4424) | (0.0514) | (0.5321) | (0.0322) | (1.2808) | (0.0102) |
| *GMPU* | -1.5014*** | -0.1750*** | -0.8002** | -0.0518** | 0.7095 | 0.0051 |
| | (0.3067) | (0.0331) | (0.3717) | (0.0233) | (0.8671) | (0.0061) |
| *GPRHI* | -11.9441*** | -1.3924*** | 3.0687 | 0.1988 | 3.2588 | 0.0232 |
| | (4.3151) | (0.4691) | (3.1653) | (0.2059) | (7.3366) | (0.0528) |
| constant | 6.9279*** | | 10.4468*** | | 6.3200 | |
| | (2.2997) | | (2.7440) | | (5.9109) | |
| Log likelihood | -191.2089 | | -126.9966 | | -28.3799 | |
| LR Statistics | 41.4000 | | 23.4100 | | 7.3700 | |
| Prob>Chi square | 0.0000 | | 0.0000 | | 0.0611 | |
| Observations | 465 | | 465 | | 465 | |
| McFadden's $R^2$ | 0.0980 | | 0.0840 | | 0.1150 | |
| | Managed Floating Regime: 1997M8-2023M9 | | | | | |
| *GPRH* | 0.7336 | 0.1833 | -1.5325* | -0.2043* | | |
| | (0.6170) | (0.1543) | (0.8205) | (0.1085) | | |
| *GMPU* | -1.5073*** | -0.3768** | -1.7749*** | -0.2366*** | | |
| | (0.5075) | (0.1269) | (0.6762) | (0.0844) | | |
| *GPRHI* | -6.3868 | -1.5966 | 6.3885 | 0.8515 | | |
| | (6.8066) | (1.7016) | (8.4389) | (1.1252) | | |
| constant | 3.5732 | | 12.6870*** | | | |
| | (2.7822) | | (3.9394) | | | |
| Log likelihood | -99.2711 | | -65.4950 | | | |
| LR Statistics | 10.7800 | | 16.7400 | | | |
| Prob>Chi square | 0.0130 | | 0.0000 | | | |
| Observations | 151 | | 151 | | | |
| McFadden's $R^2$ | 0.0520 | | 0.1130 | | | |
| | Free Floating Regime: 1997M8-2023M9 | | | | | |
| *GPRH* | 1.1289 | 0.0103 | 0.1158 | 0.0046 | 0.4650 | 0.0035 |
| | (1.6271) | (0.0147) | (0.8047) | (0.0318) | (1.8985) | (0.0144) |
| *GMPU* | 1.1356 | 0.0103 | 1.4663*** | 0.0578** | 0.8939 | 0.0068 |
| | (0.8757) | (0.0094) | (0.5555) | (0.0194) | (1.0719) | (0.0087) |
| *GPRHI* | -29.6663 | -0.2703 | -1.3041 | -0.0514 | -23.6793 | -0.1800 |
| | (18.6152) | (0.1302) | (5.3885) | (0.2122) | (20.8782) | (0.1167) |
| constant | -13.4046* | | -10.6360*** | | -9.8795 | |
| | (7.8927) | | (4.0211) | | (9.2543) | |
| Log likelihood | -26.5827 | | -59.3936 | | -19.9807 | |
| LR Statistics | 6.2100 | | 7.6400 | | 2.8900 | |
| Prob>Chi square | 0.1010 | | 0.0540 | | 0.4086 | |
| Observations | 314 | | 314 | | 314 | |
| McFadden's $R^2$ | 0.1050 | | 0.0600 | | 0.0670 | |

Note: ***, **, * are significant at 1%, 5%, and 10%, respectively.
Source: Authors' calculation

The robustness check results ratify the baseline findings in Table 2. In complete sample data, we strongly confirm that the global economic policy uncertainty negatively shakes speculative bubbles in the exchange rate and the ratio of exchange rates for non-traded. The country's geopolitical risks negatively discover explosive activities in the exchange rate only. We also approve the heterogeneity results for different exchange rate systems. In the managing float system, global- geopolitical risk and economic policy uncertainty shake the ratio of exchange rates for non-traded only, whilst, in the free-floating regime, the Indonesia-specific geopolitical risk leads to diminishing bubbles in the exchange rate and the ratio of exchange rates for traded goods components.

## 6. Conclusion

Major uncertainty trinity events can lead to explosive actions in the exchange rate. This analysis argues the first effort to feature the effect of the uncertainty trinity on explosive bubbles in the Indonesian Rupiah against the US exchange rate. Firstly, we implement the GSADF to uncover explosive bubbles and propose the uncertainty trinity as the driver of explosive behavior. A detailed set of empirical findings confirms that the GSADF detects that the Indonesian Rupiah/US exchange rate deviates from the primary values six times throughout the duration from January 1985 to September 2023, signifying the incidence of numerous explosive behaviors. When the full-sample period splits into the managed-floating regime and the free-floating regime, the GSADF still discovers multiple bubbles. Secondly, we employ logistic estimation to capture the role of the uncertainty trinity on multiple bubble episodes in the Rupiah against the US dollar.

Of particular inquisitiveness on uncertainty trinity, this study accentuates that global geopolitical risk negatively determines explosive bubbles in the ratio of exchange rates for non-traded and traded goods. In full sample data, we show that the global economic policy uncertainty negatively shakes speculative bubbles in the exchange rate and the ratio of exchange rates for non-traded. The country's geopolitical risks negatively discover explosive bubbles in the exchange rate only. Interestingly, examining different exchange rate systems, we find heterogeneity in our results. In the managing float system, global- geopolitical risk and economic policy uncertainty upset the ratio of exchange rates for non-traded only. Yet, in the free-floating regime, the Indonesia-specific geopolitical risk leads to a weakening in the exchange rate and the ratio of exchange rates for traded goods. The robustness checks further strongly establish the systematic empirical results. Thus, to stand the adverse economic magnitudes of explosive bubbles, regulators have a duty to design proper preventive policies in the face of a confounded uncertainty-trinity landscape and a volatile foreign exchange market ecosystem.

The limitations of this study are that it only focuses on the Indonesia Rupiah against the US exchange rate to detect bubble behaviors and the effect of uncertainty trinity. Yet, the position exchange rate of major strategic Indonesian partner countries in future international trading targets cannot be neglected. Consequently, the future analysis can take into account various Indonesian rupiahs against foremost trading partners' currencies in the future. The comparative explorations on the sources of bubble behaviors in exchange rates against different main partner countries over various uncertainties, like pandemic uncertainty, Indonesian-specific economic policy uncertainty, and uncertainty in animal spirit or herding behaviors can be carried out in future studies. Finally, the analysis can further be extended to the interconnectedness and spillover of bubble behavior in exchange rate issues.